\documentclass[twocolumn, twocolappendix]{aastex631}

\begin{document}

\title{Discovery of MgS and NaS in the Interstellar Medium and tentative detection of CaO}

\author{Marta Rey-Montejo}
\affiliation{Centro de Astrobiolog{\'i}a (CAB), CSIC-INTA, Ctra. de Ajalvir km 4, E--28850, Torrej\'on de Ardoz, Spain}

\author{Izaskun Jim\'enez-Serra}
\affiliation{Centro de Astrobiolog{\'i}a (CAB), CSIC-INTA, Ctra. de Ajalvir km 4, E--28850, Torrej\'on de Ardoz, Spain}

\author{Jes\'us Mart{\'i}n-Pintado}
\affiliation{Centro de Astrobiolog{\'i}a (CAB), CSIC-INTA, Ctra. de Ajalvir km 4, E--28850, Torrej\'on de Ardoz, Spain}

\author{V{\'i}ctor M. Rivilla}
\affiliation{Centro de Astrobiolog{\'i}a (CAB), CSIC-INTA, Ctra. de Ajalvir km 4, E--28850, Torrej\'on de Ardoz, Spain}

\author{Andr\'es Meg{\'i}as}
\affiliation{Centro de Astrobiolog{\'i}a (CAB), CSIC-INTA, Ctra. de Ajalvir km 4, E--28850, Torrej\'on de Ardoz, Spain}

\author{David San Andr\'es}
\affiliation{Centro de Astrobiolog{\'i}a (CAB), CSIC-INTA, Ctra. de Ajalvir km 4, E--28850, Torrej\'on de Ardoz, Spain}

\author{Miguel Sanz-Novo}
\affiliation{Centro de Astrobiolog{\'i}a (CAB), CSIC-INTA, Ctra. de Ajalvir km 4, E--28850, Torrej\'on de Ardoz, Spain}

\author{Laura Colzi}
\affiliation{Centro de Astrobiolog{\'i}a (CAB), CSIC-INTA, Ctra. de Ajalvir km 4, E--28850, Torrej\'on de Ardoz, Spain}

\author{Shaoshan Zeng}
\affiliation{Star and Planet Formation Laboratory, Cluster for Pioneering Research, RIKEN, 2–1 Hirosawa, Wako, Saitama, 351–0198, Japan}

\author{\'Alvaro L\'opez-Gallifa}
\affiliation{Centro de Astrobiolog{\'i}a (CAB), CSIC-INTA, Ctra. de Ajalvir km 4, E--28850, Torrej\'on de Ardoz, Spain}

\author{Antonio Mart{\'i}nez-Henares}
\affiliation{Centro de Astrobiolog{\'i}a (CAB), CSIC-INTA, Ctra. de Ajalvir km 4, E--28850, Torrej\'on de Ardoz, Spain}

\author{Sergio Mart{\'i}n}
\affiliation{European Southern Observatory, Alonso de C\'ordova, 3107, Vitacura, Santiago 763-0355, Chile}
\affiliation{Joint ALMA Observatory, Alonso de C\'ordova, 3107, Vitacura, Santiago 763-0355, Chile}

\author{Bel\'en Tercero}
\affiliation{Observatorio Astron\'omico Nacional (OAN-IGN), Calle Alfonso XII, 3, 28014 Madrid, Spain}
\affiliation{Observatorio de Yebes (OY-IGN), Cerro de la Palera SN, Yebes, Guadalajara, Spain}

\author{Pablo de Vicente}
\affiliation{Observatorio de Yebes (OY-IGN), Cerro de la Palera SN, Yebes, Guadalajara, Spain}

\author{Miguel Angel Requena Torres}
\affiliation{Department of Physics, Astronomy, and Geosciences, Towson University, Towson, MD 21252, USA}

\begin{abstract}

We report the first detection of the metal-bearing molecules sodium sulfide (NaS) and magnesium sulfide (MgS) and the tentative detection of calcium monoxide (CaO) in the interstellar medium (ISM) towards the Galactic Center molecular cloud G+0.693-0.027. The derived column densities are $(5.0\pm1.1)\times10^{10}$ cm$^{-2}$, $(6.0\pm0.6)\times 10^{10}$ cm$^{-2}$, and $(2.0\pm0.5)\times10^{10}$ cm$^{-2}$, respectively. This translates into fractional abundances with respect to H$_2$ of $(3.7\pm1.0)\times10^{-13}$, $(4.4\pm0.8)\times10^{-13}$, and $(1.5\pm0.4)\times10^{-13}$, respectively. We have also searched for other Na-, Mg- and Ca-bearing species towards this source but none of them have been detected and thus we provide upper limits for their abundances. We discuss the possible chemical routes involved in the formation of these molecules containing metals under interstellar conditions. Finally, we compare the ratio between sulfur-bearing and oxygen-bearing molecules with and without metals, finding that metal-bearing sulfur molecules are much more abundant than metal-bearing oxygen ones, in contrast with the general trend found in the ratios between other non metal- oxygen- and sulfur-bearing molecules. This further strengthen the idea that sulfur may be little depleted in G+0.693-0.027 as a result of the low velocity shocks present in this source sputtering large amounts of material from dust grains.

\end{abstract}

\keywords{Molecular clouds (1072) --- Galactic center(565) --- Interstellar molecules(849) --- Astrochemistry(75) --- Spectral line identification(2073)}

\section{Introduction} \label{sec:intro}

Metals\footnote{We refer to metals as defined in chemistry, i.e. as every substance with high electrical conductivity, luster, and malleability, which readily loses electrons to form positive ions. These include alkali metals such as Li, Na, K, Rb, Cs and Fr, and alkaline earth metals such as Be, Mg, Ca, Sr, Ba and Ra \citep{atkins2018}.} constitute a significant component of the interstellar medium (ISM) and they are known to be heavily depleted onto dust grains in dense molecular clouds \citep{Field1974, Savage1996, Savaglio2003, Jenkins2009, DeCia2016, Roman2021, Konstant2023, Konstantopoulou2024}. These dust grains are mainly composed by graphites and silicates,
\citep{LaorDraine1993, Jaeger1994, Dorschner1995, Mutschke1998, TIELENS1998, Fabian2001}, with differences in the chemical composition depending on the type of astronomical source \citep{koheler2010}.
Graphites are mainly composed of carbon \citep{Dartois2019}. However, silicates present heavier metal elements. We typically consider seven types of silicates: olivine (whose composition is Mg$_{2x}$Fe$_{2-2x}$SiO$_4$), forsterite (Mg$_2$SiO$_4$), fayalyte (Fe$_2$SiO$_4$), pyroxene\footnote{Pyroxene could also present a sodium/aluminum composition of the form Na$_{0.5}$Al$_{0.5}$SiO$_3$ \citep{min2007}.} (Mg$_x$Fe$_{1-x}$SiO$_3$), hematite (Fe$_2$O$_3$), gehlenite (Ca$_2$Al$_2$SiO$_7$), and enstatite (MgSiO$_3$). 

Dust grains are also believed to be an important reservoir of sulfur, since the abundance of sulfur volatile species such as H$_2$S, SO and SO$_2$ in molecular clouds is orders of magnitude lower than the sulfur cosmic abundance; this is known as the {\it sulfur depletion problem} \citep[][]{agundez2013}. Sulfide-minerals, such as FeS nano-inclusions, have indeed been found in silicate dust \citep{Koheler2014} and in asteroids \citep{matsumoto2020}. These inclusions could deplete one-third of the cosmic sulfur into interestellar dust without inducing a significant change in the observed mid-IR silicate absorption bands \citep{Koheler2014}. In addition, in protoplanetary disks sulfide-minerals could account for almost 90\% of the elemental sulfur depleted in these objects \citep{kama2019}. However, metal-sulfide molecules such as NaS or MgS have not been detected in the ISM so far, and thus, it is not possible to estimate in what amount they contribute to the sulfur depletion problem.  

Since metals are highly refractory, metal-bearing molecules are difficult to detect in the gas phase. As a result, this type of molecular species have been reported only towards the molecular envelopes of evolved stars and a few star-forming regions. 
Most of the detections of metal-bearing molecules have been achieved towards the circumstellar medium (CSM) of the evolved star IRC+10216. \citet{CernicharoGuelin1987} reported the first detections of diatomic halides such as NaCl, AlCl, KCl and AlF towards this object. Metals in the form of neutral atoms and ions seem to survive in the gas phase in the outer envelope of the evolved star, which suggests a rich metal chemistry in this cooler region of the envelope \citep{MauronHuggings2010}. 

After these early detections, other metal cyanides and isocyanides, and metal-containing carbon chains have also been reported towards the same source \citep[see e.g. MgNC, NaCN, MgCN, AlNC, KCN, FeCN, HMgNC, MgCCH, CaNC, MgC$_3$N, MgC$_4$H, MgC$_5$N, MgC$_6$H, MgC$_2$, MgC$_4$H$^+$, MgC$_6$H$^+$, MgC$_3$N$^+$, MgC$_5$$^+$, HMgCCCN and NaCCCN;][]{Kawaguchi1993, Guelin1993, Turner1994, Ziurys1995, Ziurys2002, Pulliam2010, Zack2011, Cabezas2013, Agundez2014, Cernicharo2019a, Cernicharo2019b, Pardo2021, Changala2022, Cernicharo2023, Cabezas2023}. NaCl has also been observed towards the ejecta of other evolved stars such as IK Tauri and VY Canis Majoris \citep{Milam2007, Kaminski2013}, CRL 2688 \citep{Highberger2003b} and OH 231+4.2 \citep{Sanchez2018}. 

In the ISM, metal-bearing molecules have been detected towards three sources, the disk around the Orion source I \citep[like NaCl, KCl and AlO;][]{Ginsburg19, Tachibana19}, the disk around high-mass young stellar objects \citep[NaCl and KCl;][]{gingsburg2023}, and the IRAS 16547-4247 binary system \citep[only NaCl;][]{Tanaka2021}. Regarding Na-bearing molecules, upper limits have been derived in diffuse clouds, such as NaH towards $\zeta$ Oph and $\zeta$ Per \citep{Czarny1987}.
Despite exhaustive searches of MgO \citep{Turner1985}, MgS \citep{Takano1989} and MgH \citep{Czarny1987}, Mg-bearing molecules remain undetected in the gas phase in the ISM, with upper limits for $N(\rm MgH)<1.9\times10^9$ cm$^{-2}$ towards $\zeta$ Oph. The same scenario applies to Ca-containing molecules such as CaO \citep{Hocking1979} and CaS \citep{Takano1989}, with upper limits lower than $5-6\times10^{11}$ cm$^{-2}$ toward a sample of high-mass star-forming regions. 

In this paper we present the discovery of the metal-bearing molecules NaS and MgS, and the tentative detection of CaO in the ISM. 
These detections have been achieved 
towards the Galactic Center molecular cloud G+0.693-0.027 (hereafter G+0.693) which presents a very rich chemistry due to large-scale shocks likely induced by a cloud-cloud collision \citep{zeng2020}. These shocks sputter significant amounts of the material contained within the icy mantles and refractory cores of dust grains, releasing many molecular species into the gas phase. In the past few years, spectral surveys carried out towards G+0.693 have provided the detection of more than a dozen molecules towards this source
\citep[see, e.g.,][]{jimenez2020, Jimenez2022, Rivilla2019, Rivilla2020, Rivilla2021a, Rivilla2021b, Rivilla2022a, Rivilla2022b, Rivilla2023, Rodriguez2021a, Rodriguez2021b, sanznovo2023, Sanz2024, Sanz2024b, Zeng2021, Zeng2023, david2024}, which places G+0.693 as one of the best astronomical objects for discovering new interstellar species in the gas phase, including refractory molecular material.

The paper is organized as follows. Section~\ref{sec:Obs} describes the observations carried out towards G+0.693. In section~\ref{sec:Analysis and Results}, we 
analyse the emission from all metal-bearing molecules towards this source and present the LTE analysis of the detections of MgS, NaS and CaO. We also report the upper limits derived for other Na-, Mg- and Ca-bearing molecules, which remain undetected. In section~\ref{sec:Discussion} we compare the derived column densities and upper limits with those measured towards other astronomical sources and discuss the main findings of our study. Finally, Section~\ref{sec:Conc} collects our conclusions.

\section{Observations} \label{sec:Obs}

We have analysed the new ultrasensitive, broadband spectral survey towards G+0.693 \citep{Rivilla2023, sanznovo2023}. These observations were carried out with the Yebes 40 m (Guadalajara, Spain) IRAM 30 m (Granada, Spain) and APEX 12 m (Atacama, Chile) radiotelescopes in position-switching mode. As central coordinates, we used  $\alpha = 17^{\rm h}47^{\rm m}22^{\rm s}$, $\delta=-28^{\circ} 21\arcmin 27\arcsec$, with the off position shifted by $\Delta\alpha=-885\arcsec$ and $\Delta\delta=290\arcsec$. The location of G+0.693 corresponds to source Sgr B2M (20$"$, 100$"$) in Table$\,$1 of \citet{Martin2008}.

For the Yebes 40 m observations, the ultra-broadband Nanocosmos Q-band (7 mm) HEMT receiver was used to cover the whole frequency range (from 31.07 to 50.42 GHz) in two linear polarizations, providing a channel witdth of 38 kHz \citep{Tercero2021}. The IRAM 30 m observations were carried out using the Eight MIxer Receivers (EMIR) and the Fast Fourier Transform Spectrometer (FTS200) providing a spectral resolution of 195 kHz along the three frequency ranges: 83.2-115.41, 132.28-140.39, and 142-173.81 GHz. For the APEX observations the NFLASH230 receiver was connected to two FFTS backends, which provided a simultaneous coverage of two sidebands of 7.9 GHz each separated by 8 GHz. The spectral resolution was 250 kHz. The covered spectral ranges were 217.9-225.9, 234.18-242.18, 243.94-252.12 and 260.18-268.37 GHz. 

The spectra of Yebes and IRAM observations were finally smoothed to a frequency resolution of 256 kHz (1.5-2.5 km s$^{-1}$) and 615 kHz (1.0-2.2 km s$^{-1}$), respectively, while for the APEX data, the final spectra were smoothed to 1 MHz or 1.1-1.4 km s$^{-1}$. 
The half power beam width (HPBW) was $\sim$ 35$\arcsec$-55$\arcsec$ for Yebes 40m, $\sim$ 14$\arcsec$-29$\arcsec$ for IRAM 30m, and $\sim$23$\arcsec$-28$\arcsec$ for APEX in the observed frequency ranges. Note, however, that the molecular emission towards G+0.693 is extended over the beam \citep{Requena2006, Requena2008, Zeng2018, zeng2020} and hence, the measured spectra are given in units of antenna temperature (T$_A^*$). We refer to \citet{Rivilla2022b,Rivilla2023} and \citet{sanznovo2023} for further information on all observations and data reduction. For the frequency ranges not covered by these new observations, we used the data from our previous IRAM 30m survey \citep[see][]{Rodriguez2021a, Rivilla2022b}.

\section{Analysis and Results} \label{sec:Analysis and Results}

We have used the line survey towards G+0.693 to search for metal-bearing molecules for which rotational spectroscopy is available (see Table~\ref{tab:spec}).
Our observations has allowed us to detect two sulfur-bearing molecules, sodium sulfide (NaS) and magnesium sulfide (MgS), towards G+0.693. We also report the tentative detection of calcium oxide (CaO) for the first time in the ISM (physical parameters in Table~\ref{tab:PP}). For the rest of metal-bearing molecules searched for in our dataset, we only obtained upper limits (see Table~\ref{tab:uplim}). 

The identification of the molecular lines was carried out using the Spectral Line Identification and Modeling (\textsc{Slim}) tool from \textsc{Madcuba}, which assumes local thermodynamic equilibrium (LTE) excitation to compute the synthetic spectra. To derive the physical parameters of the emission, we used the \textsc{Slim}-\textsc{Autofit} tool of \textsc{Madcuba} \citep{Martin2019} that provides the best nonlinear least-squares LTE fit to the data using the Levenberg-Marquardt algorithm. The fit provides information about the column density, $N$, excitation temperature, $T_{\rm ex}$, radial velocity, $v_{\rm LSR}$, and linewidth, FWHM, of the molecular emission. In the following, we present the results of the analysis of the different chemical families using \textsc{Madcuba}-\textsc{Slim}.

\begin{deluxetable}{ccccc} 
\tablecaption{Physical parameters of the detected molecules towards G+0.693.}
\tablehead{\colhead{Molecule}  & \colhead{$N$} & \colhead{$T_{\rm ex}$} & \colhead{$v_{\rm LSR}$} & \colhead{FWHM}
\\
\colhead{} & \colhead{($10^{10}$ cm$^{-2}$)} & \colhead{(K)} & \colhead{(km s$^{-1}$)} & \colhead{(km s$^{-1}$)}
}
\startdata 
NaS & 5.0$\pm$1.1 & 17.5$\pm$3.7 & $66.5\pm1.5$ & 15$^a$\\
MgS & 6.0$\pm$0.6 & 37.9$\pm$4.8 & 69.0$^a$ & 15$^a$\\
CaO & 2.0$\pm$0.5 & 28.6$\pm$9.6 & 69.0$^a$ & 20$^a$\\
\enddata
\tablenotetext{a}{Fixed parameter.}
\label{tab:PP}
\end{deluxetable}

\begin{figure*}
    \centering
\plotone{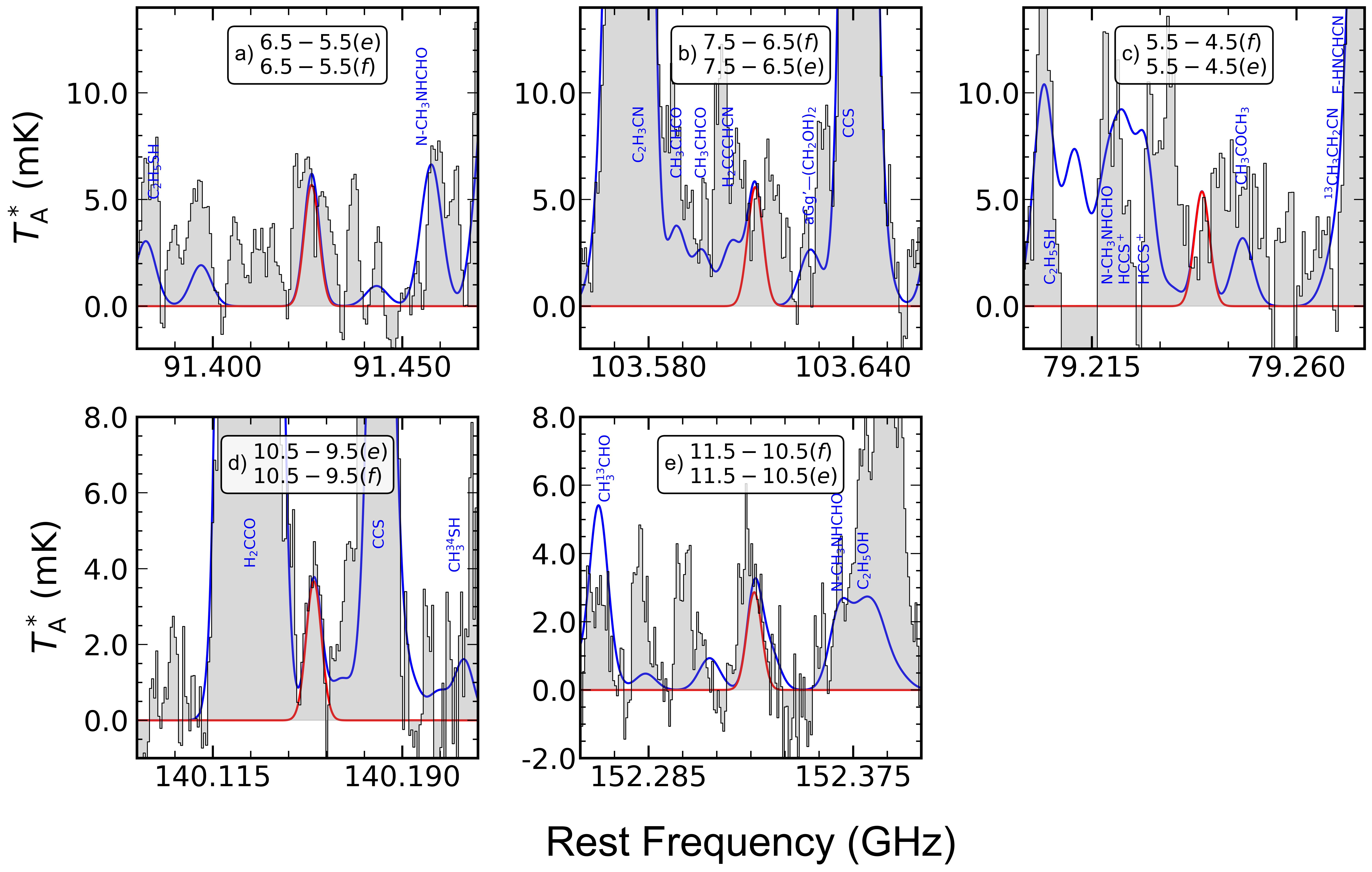}
\caption{Unblended transitions of sodium sulfide (NaS) detected towards G+0.693, displayed in order of decreasing line intensity. The observed spectrum is represented in grey. The best LTE fit derived with \textsc{Madcuba} is shown with a red line, while the blue line shows the expected molecular emission from all of the molecular species identified to date in our survey. The quantum numbers of each transition are shown in the upper part of each panel, and indicate the $J$+1$\rightarrow$$J$ transition and the parity (e or f). All the transitions shown here correspond to the sub-state $\Pi^2_{3/2}$ (see text). 
\label{fig:NaS}}
\end{figure*}

\subsection{Detection of sodium sulfide (NaS)}
 \label{sec:Nas}
Nine rotational transitions of sodium sulfide (NaS) have been covered within our spectroscopic survey towards G+0.693. Five out of nine transitions appear unblended from known molecular species, while the remaining four are either heavily blended with more abundant species previously identified towards G+0.693 or fall within the noise. For this analysis, we have thoroughly searched all the molecules included in the CDMS \citep{endres2016}, JPL \citep{pickett1998}, and Lille\footnote{See https://lsd.univ-lille.fr/.} spectroscopic databases in the vicinity of the NaS target lines, and no transitions from other molecular species besides NaS are able to explain the observed spectral features. Despite this, the transitions 15/2$\rightarrow$13/2 and 11/2$\rightarrow$9/2 appear blended with unidentified line emission (U-line). However, as explained in Section \ref{sec:validity}, the level of blending is $\leq$25\% and therefore, we consider them as ``Unblended$"$. 

In Figure~\ref{fig:NaS} and Table~\ref{tab:NaS}, we present the observed spectra and the spectroscopic information of the three unblended transitions of NaS, together with the two lines blended with U-line emission. In Table~\ref{tab:NaS}, we show that each observed line is produced by two different lambda states of the same transition, which are labeled by the parity notation ``e$"$ and ``f$"$. Furthermore, all transitions detected toward G+0.693 correspond to the $\Pi^2_{3/2}$ state. This is because the ground state of NaS is inverted, i.e. the $\Omega$ = 3/2 component lies lower in energy than the $\Omega$ = 1/2 sublevel \citep{Li1997}. This explains why the transitions of NaS in the $\Omega$ = 3/2 sublevel are easier to detect
than the ones in the $\Omega$ = 1/2 sublevel. 
The big absorption feature seen at 79.238424-79.238730 GHz in the third panel of Figure~\ref{fig:NaS}, may be due to a molecular transition affected by strong non-LTE effects, or to emission in the OFF position used in our observations. 

We note that three unblended transitions are enough to claim the detection of metal-bearing molecules (see the validity of the NaS detection in Section \ref{sec:validity}). As other examples, we refer to the recent discovery of FeC and SiP in the complex and line rich spectra of the evolved star IRC+10216 \citep[][]{koelemay22,koelemay23}, with respectively three and four rotational transitions reported.
In addition, the NaS transitions shown in Figure~\ref{fig:NaS} and Table~\ref{tab:NaS}, correspond to the $J$+1$\rightarrow$$J$ progression of NaS rotational lines that was covered in our broadband observations across (non-continuous) 124 GHz, and that appear unblended (or blended with U-emission) according to our criteria (see Section \ref{sec:validity}). Because of all this, the data presented here give very strong evidence in support of the presence of NaS in the ISM.

The physical parameters of the emission of NaS are reported in Table~\ref{tab:PP} and were derived using \textsc{Madcuba}-\textsc{Autofit} on the unblended transitions shown in Figure~\ref{fig:NaS} that take into account the contribution of all the molecules already identified and fitted in the molecular cloud to ensure that the observed spectral features correspond to sodium sulfide.   
To allow the convergence of the \textsc{Autofit}, we fixed the FWHM value to 15 km s$^{-1}$, which reproduces well the profile of the unblended lines, and left the molecular column density ($N$), the excitation temperature ($T_{\rm ex}$) and the central radial velocity ($v_{\rm LSR}$) as free parameters. Note that the molecular line emission measured towards G+0.693 presents typical linewidths of $\sim$15-20 km s$^{-1}$ \citep[see][]{Requena2006,Zeng2018,jimenez2020,Rivilla2020}.
The best LTE fit is shown in Figure~\ref{fig:NaS} and provides a velocity of $v_{\rm LSR}=66.5\pm1.5$ km s$^{-1}$, which is also consistent with the values found in G+0.693 for other molecular species. The excitation temperature is $T_{\rm ex}=17.5\pm3.7$ K, while the derived column density is $N=(5.0\pm1.1)\times10^{10}$ cm$^{-2}$. Assuming a H$_2$ column density of $N_{\rm H_2}=1.35\times10^{23}$ cm$^{-2}$ inferred from C$^{18}$O emission at the velocity component of G+0.693 \citep[see][]{Martin2008}\footnote{The uncertainty in the H$_2$ column density toward G+0.693 is assumed to be of 15\% its value. However note that this estimate may suffer from an additional bias by a factor of 2 because \cite{Martin2008} used the relationship C$^{18}$O/$N_{\rm H_2} =1.7\times10^{-7}$ cm$^{-2}$ \citep{Frerking1982} derived from the molecular clouds in the Galactic disk, and there is a factor of 2 difference in the isotopic ratio $^{16}$O/$^{18}$O measured in the Galactic Center compared to the Solar neighbourhood. This means that the abundance of NaS could be a factor of 2 higher.}, the NaS column density translates into a molecular abundance with respect to H$_2$ of $(3.7\pm1.0)\times10^{-13}$ towards G+0.693. However, this value should be considered as a lower limit since NaS may be produced in a region smaller than that probed by C$^{18}$O (see a detailed discussion in Section \ref{sec:uncertainties}).

\begin{deluxetable*}{cccccccccc} 
\tablecaption{List of observed transitions of NaS in order of decreasing line intensity.}
\tablehead{\colhead{Frequency} & \multicolumn{3}{c}{Transition} & \colhead{log I} & \colhead{E$_{\rm u}$}  &  \colhead{rms} & \colhead{$\int T_A ^* \rm dv$$^d$} & \colhead{S/N$^a$} & \colhead{Blending}
\\
\colhead{(GHz)} & \multicolumn{3}{c}{($J\rightarrow J'$ \ Parity \ $\Pi_i ^2$)} & \colhead{($\rm nm^{2}~MHz$)} & \colhead{(K)} & \colhead{(mK)} & \colhead{(mK km s$^{-1}$)} & 
\colhead{} & \colhead{}
}
\startdata 
91.425124 (80)$^b$ & 13/2 $\rightarrow$ 11/2 & e & $\Pi^2_{3/2}$ & -1.9179 & 17.5 & 1.7 & 92 (9)$^c$ & 10.0 & Unblended*\\
91.425530 (80)$^b$ & 13/2 $\rightarrow$ 11/2 & f & $\Pi^2_{3/2}$ & -1.9179 & 17.5 &  &  &  & Unblended*\\
103.610186 (86)$^b$ & 15/2 $\rightarrow$ 13/2, & f & $\Pi^2_{3/2}$ & -1.7576 & 22.4 & 2.2 & 90 (12)$^c$ & 7.8 & Blended with U-lines*\\
103.610709 (86)$^b$ & 15/2 $\rightarrow$ 13/2, & e & $\Pi^2_{3/2}$ & -1.7576 & 22.4 &  &  &  & Blended with U-lines*\\
79.238424 (72)$^b$ & 11/2 $\rightarrow$ 9/2, & f & $\Pi^2_{3/2}$ & -2.1044 & 13.1 & 1.9 & 86 (11)$^c$ & 7.6 & Blended with U-lines*\\
79.238730 (72)$^b$ & 11/2 $\rightarrow$ 9/2, & e & $\Pi^2_{3/2}$ & -2.1044 & 13.1 &  &  &  & Blended with U-lines*\\
140.153378 (97)$^b$ & 21/2 $\rightarrow$ 19/2, & e & $\Pi^2_{3/2}$ & -1.3831 & 40.7 & 1.4 & 60 (6)$^c$ & 10.0 & Unblended*\\
140.154336 (97)$^b$ & 21/2 $\rightarrow$ 19/2, & f & $\Pi^2_{3/2}$ & -1.3831 & 40.7 &  &  &  & Unblended*\\
152.329716 (97)$^b$ & 23/2 $\rightarrow$ 21/2, & f & $\Pi^2_{3/2}$ & -1.2835 & 48.0 & 1.0 & 46 (4)$^c$ & 10.6 & Unblended*\\
152.330848 (97)$^b$ & 23/2 $\rightarrow$ 21/2, & e & $\Pi^2_{3/2}$ & -1.2835 & 48.0 &  &  &  & Unblended*\\
\enddata
\tablecomments{We provide the transition frequencies, quantum numbers, base 10 logarithms of the integrated intensity at 300 K (log I), energies of the upper levels (E$_{\rm u}$), noise level (rms), integrated area ($\int T_A ^* dv$), and signal to noise ratio (S/N) of each transition. The last column indicates the line blending with other molecular species. ``Unblended" lines refer to those transitions that are not contaminated by the emission from other known species. We also use an asterisk symbol for those lines that are (auto)blended with another transition of NaS.}
\tablenotetext{a}{ The S/N is calculated from the integrated signal ($\int T_A ^* \rm dv$) and the noise level $\sigma= \rm rms\times\sqrt{\delta v \times \rm FWHM}$, where $\delta v$ is the velocity resolution of the spectra, and the FWHM is fitted from the data (see Section~\ref{sec:Obs}).}
\tablenotetext{b}{Frequency uncertainty to the last two digits.}
\tablenotetext{c}{Fitted integrated line intensity uncertainty.}
\tablenotetext{d}{Fitted integrated line intensity for the two transitions conforming each line.}
\label{tab:NaS}
\end{deluxetable*}

\begin{figure}
    \centering
    \includegraphics[width=0.48\textwidth]{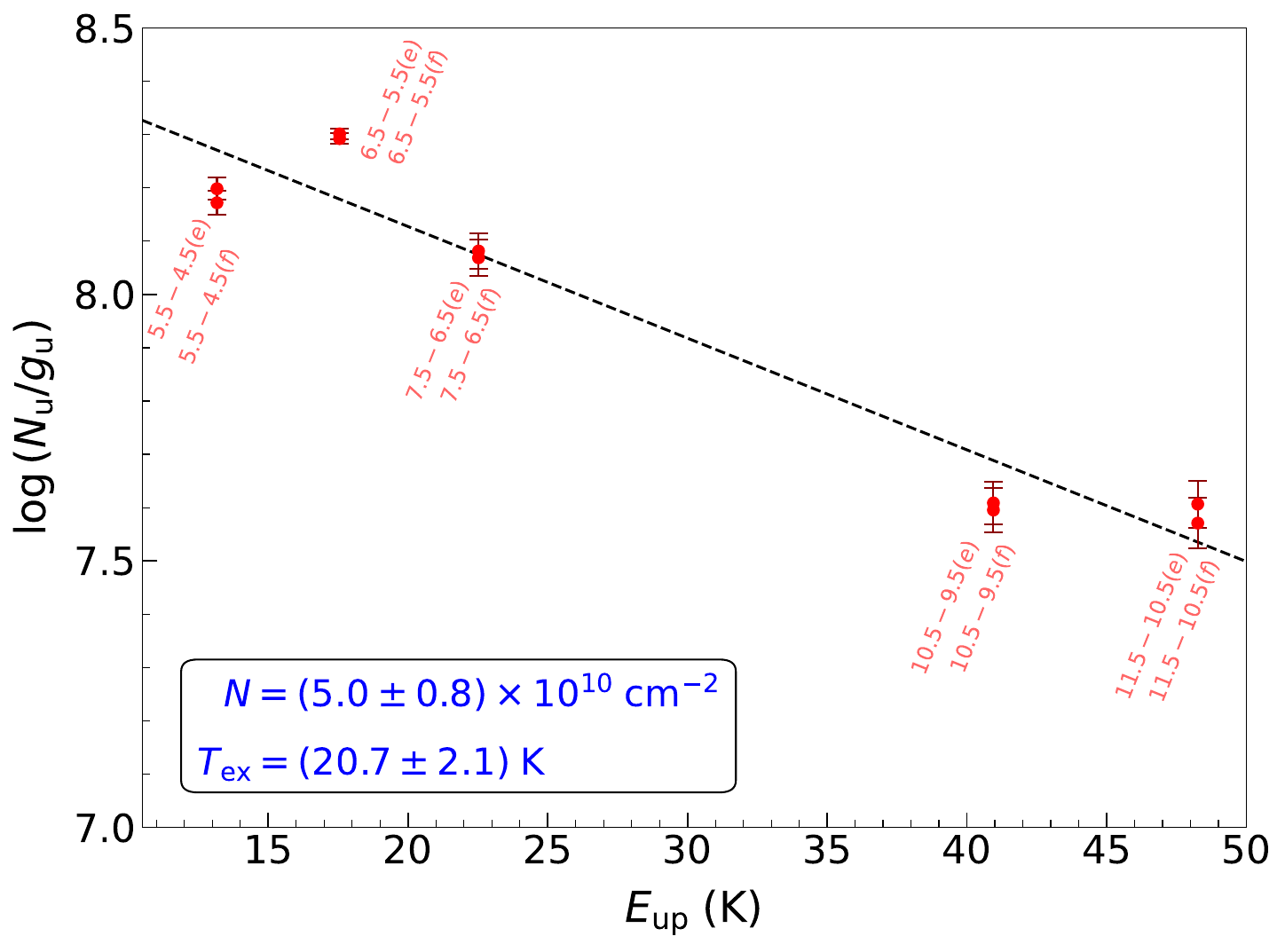}
\caption{Rotational diagram of NaS (red dots) using the transitions reported in Table~\ref{tab:NaS}. The black dashed line corresponds to the best linear fit to the data points. The derived values for the molecular column density, $N$, and the excitation temperature, $T_{\rm ex}$, are shown in blue. Ten points are represented in the figure, corresponding to the different parity of each line due to lambda doubling.}
\label{fig:Nas2}
\end{figure}

We have also carried out a rotational diagram analysis (see Figure~\ref{fig:Nas2}) of the emission of NaS with \textsc{Madcuba} considering the observed velocity-integrated intensity over the linewidth \citep{Rivilla2021a}. As described in \citet{sanandres2024}, MADCUBA’s rotational diagram considers the predicted emission from all of the molecular species previously detected in the source, and accounts for the blending in each transition by subtracting the predicted line profiles of the blending species from the observed spectrum. In this way, the rotational diagram can be derived using an ``unblended$"$ dataset of molecular transitions for the species of interest. For NaS, the detected molecular lines are basically unblended, and hence the subtraction of the blending contribution is minimal. As shown in Figure$\,$\ref{fig:Nas2}, the derived physical parameters of NaS are consistent with the ones obtained with the \textsc{Autofit} analysis: $T_{\rm ex}=20.7\pm2.1$ K and $N=(5.0\pm0.8)\times10^{10}$ cm$^{-2}$.\\

We note that the derived T$_{ex}$ is much lower than the measured kinetic temperature of the molecular gas toward this source \citep[the derived T$_{kin}$ of the molecular gas in G+0.693 lies between $\sim$70-150 K;][]{Zeng2018}. This is the usual behavior for the emission of molecules with large dipole moments in Galactic Center molecular clouds because their emission is sub-thermally excited due to the relatively low H$_2$ volume gas densities of a few 10$^4$$\,$cm$^{-3}$ \citep{Requena2006,Zeng2018,zeng2020}. However, the intensity of the molecular lines from NaS are well described by assuming a single T$_{ex}$, indicating an effect of quasi thermalization at a temperature significantly lower than T$_{kin}$, as a consequence of the low H$_2$ volume densities of the molecular gas in this source \citep{goldsmith1999}.

\subsection{Detection of magnesium sulfide (MgS)}
\label{sec:Mg}

\begin{figure*}
    \centering
\plotone{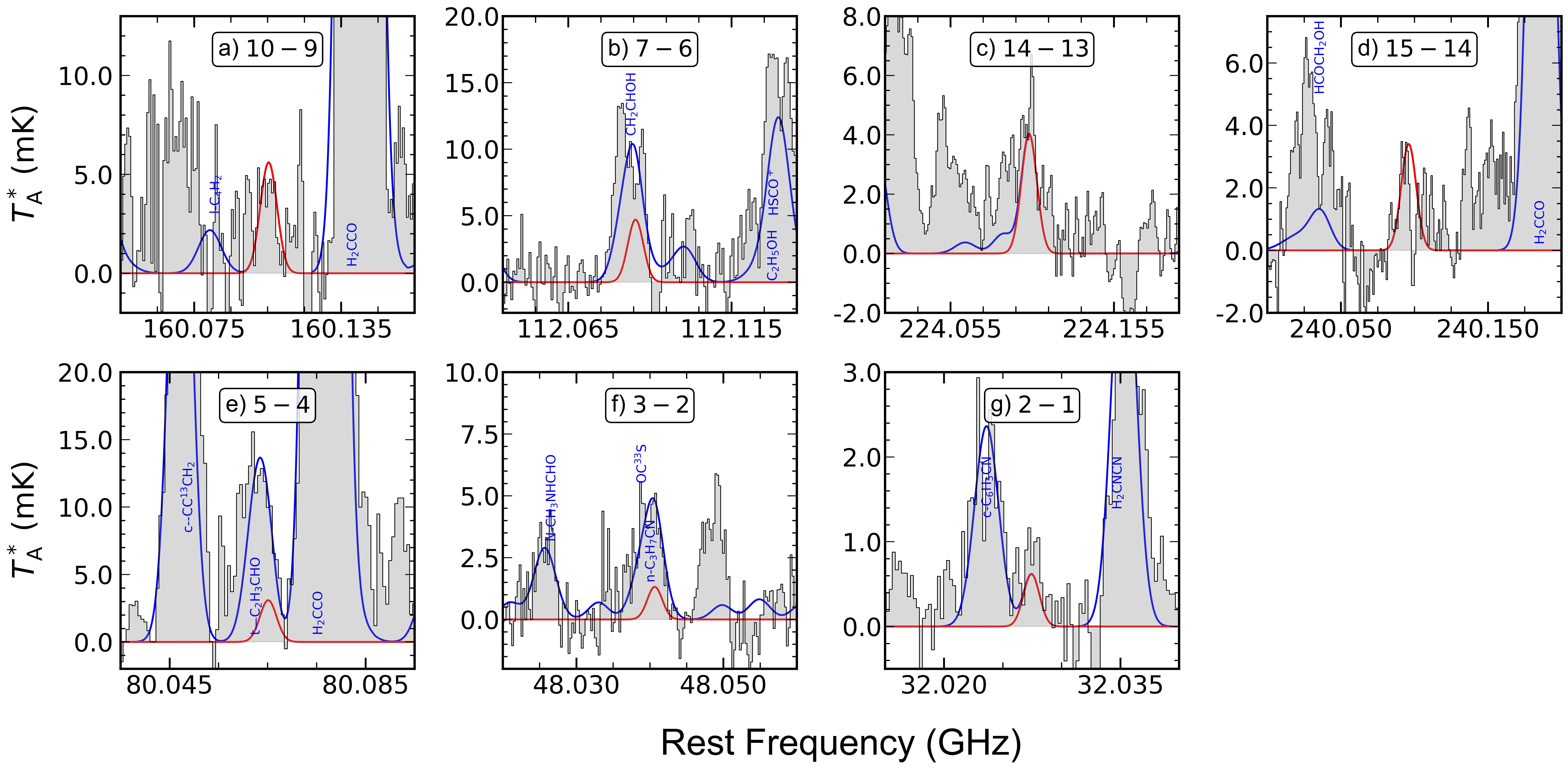}
\caption{Transitions of magnesium sulfide (MgS) detected towards G+0.693, displayed in order of decreasing line intensity. Four of the observed transitions are unblended while the other three appear blended with emission from other molecular lines. The spectra are represented as grey histograms. The best LTE fit derived with \textsc{Madcuba} is shown with a red line, while the blue line shows the predicted molecular emission from all of the molecular species identified to date in our survey. The quantum numbers of each transition are shown in the upper part of each panel. 
\label{fig:MgS}}
\end{figure*}

Following the previous methodology, we have covered up to twelve rotational transitions of magnesium sulfide (MgS) within our spectroscopic survey. Three transitions out of the 12 are unblended, while one is blended with a U-line and three other appear blended with known molecular species but reproduce almost perfectly the observed spectra (see Figure~\ref{fig:MgS} and Table~\ref{tab:MgS}). The five remaining transitions are heavily blended with other molecules towards G+0.693 or lie within the noise and are therefore not shown. 

Figure~\ref{fig:MgS} also reports the best LTE fit of MgS for which two parameters were fixed to permit the convergence of \textsc{Autofit}. The velocity was fixed to $v_{\rm LSR}=69.0$ km s$^{-1}$ and the linewidth was set to 15 km s$^{-1}$, which reproduces well the profile of the unblended lines. The derived excitation temperature is $T_{\rm ex}=37.9\pm4.8$ K, which is more than a factor of 2 higher than that inferred from the emission of NaS. In Section~\ref{sec:Discussion} we will discuss the possible origin for the difference in Tex found between NaS and MgS. The derived column density is $N=(6.0\pm0.6)\times 10^{10}$ cm$^{-2}$, (see Table~\ref{tab:PP}), which translates into a molecular abundance of $(4.4\pm0.8)\times10^{-13}$ with respect to H$_2$. 

\begin{figure}
    \centering
    \includegraphics[width=0.48\textwidth]{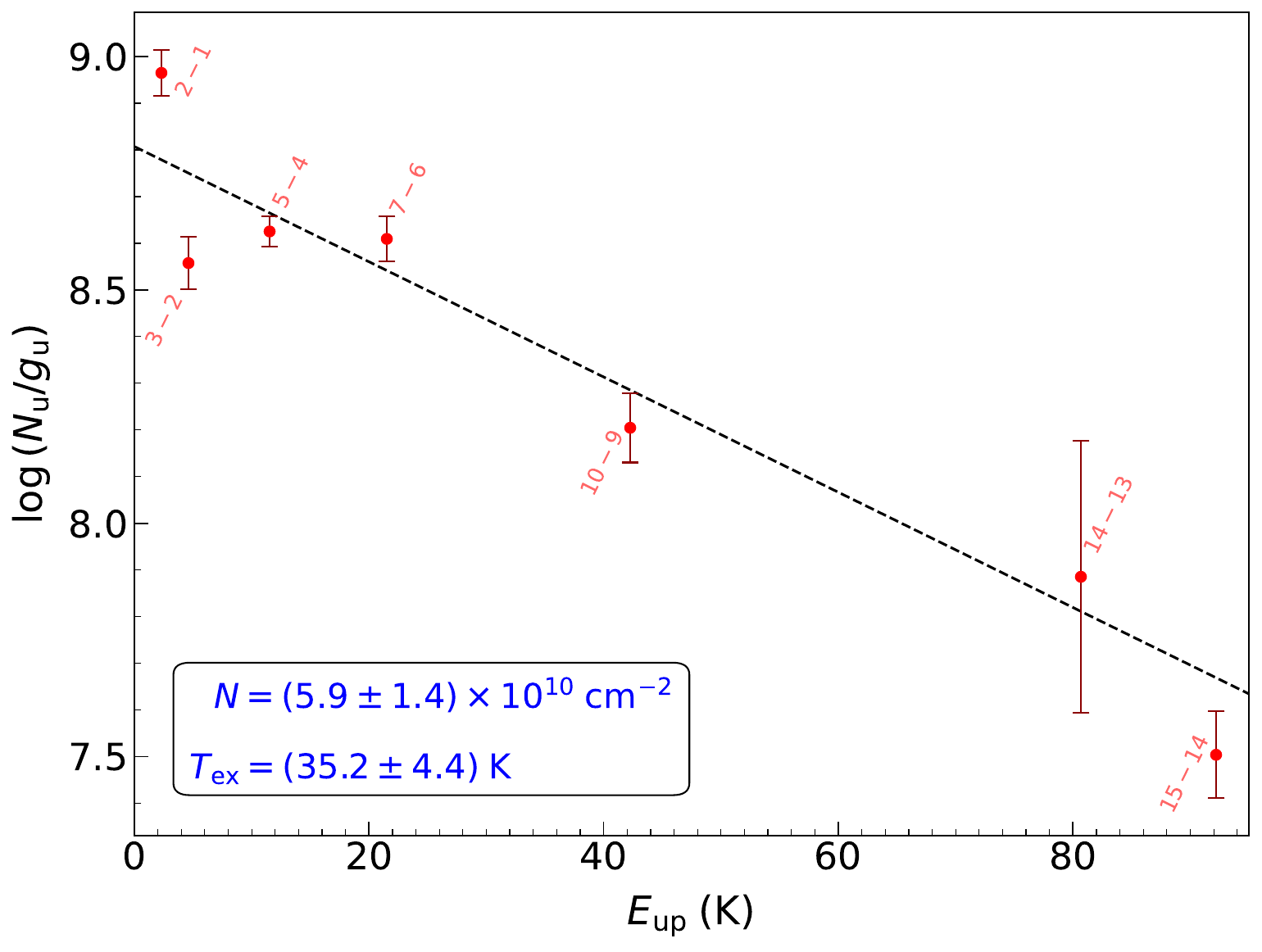}
\caption{Rotational diagram of MgS (red dots) using the transitions reported as unblended in Table~\ref{tab:MgS}. The black dashed line corresponds to the best linear fit to the data points. The derived values for the molecular column density, $N$, and the excitation temperature, $T_{\rm ex}$, are shown in blue.
\label{fig:MgSRot}}
\end{figure}

Additionally, we have carried out a rotational diagram analysis considering the velocity-integrated intensity over the linewidth \citep{Rivilla2021a}, which is shown in Figure~\ref{fig:MgSRot}. To generate this diagram with \textsc{Madcuba} \citep[see][]{david2024}, we have used the transitions reported in Figure~\ref{fig:MgS}. In this case, the rotational diagram was constructed by subtracting the contribution from any species that appears blended with the MgS transitions, as explained in Section$\,$\ref{sec:Nas}. The derived column density and excitation temperature for MgS, $N=(5.9\pm1.4)\times10^{10}$ cm$^{-2}$ and $T_{\rm ex}=35.2\pm4.4$ K, are consistent with those inferred with \textsc{Autofit}.

\begin{deluxetable*}{cccccccc} 
\tablecaption{Observed transitions of MgS towards G+0.693.}
\tablehead{\colhead{Frequency} & \colhead{Transition} & \colhead{log I} & \colhead{E$_{\rm u}$}  &  \colhead{rms} & \colhead{$\int T_A ^* \rm dv$} & \colhead{S/N$^a$} & \colhead{Blending}
\\
\colhead{(GHz)} & \colhead{($J$-$J'$)} & \colhead{($\rm
nm^{2}~MHz$)}& \colhead{(K)} & \colhead{(mK)} & \colhead{(mK km s$^{-1}$)} & 
\colhead{} & \colhead{}
}
\startdata 
160.105451 (3)$^b$ & 10-9 & -1.0416 & 42.1 & 1.9 & 108 (9)$^c$ & 12.1 & Unblended\\
112.085630 (2)$^b$ & 7-6 & -1.4779 & 21.4 & 1.2 & 90 (7)$^c$ & 13.0 & s-CH$_2$CHOH\\
224.103178 (15)$^b$ & 14-13 & -0.6568 & 80.3 & 1.0 & 79 (4)$^c$ & 20.4 & Unblended\\
240.096126 (3)$^b$ & 15-14 & -0.5831 & 91.7 & 0.8 & 67 (3)$^c$ & 20.2 & Unblended\\
80.065135 (2)$^b$ & 5-4 & -1.9029 & 11.5 & 1.7 & 59 (12)$^c$ & 5.1 & t-C$_2$H$_3$CHO\\
48.040669 (1)$^b$ & 3-2 & -2.5595 & 3.9 & 0.6 & 25 (5)$^c$ & 4.9 & OC$^{33}$S and n-C$_3$H$_7$CN\\
32.027444 (0.7)$^b$ & 2-1 & -3.085 & 2.3 & 0.3 & 12 (3)$^c$ & 3.5 & Blended with U-line\\
\enddata
\tablecomments{We provide the transition frequencies, quantum numbers, base 10 logarithm of the integrated intensities at 300 K (log I), energies of the upper levels (E$_{\rm u}$), noise level (rms), area ($\int T_A ^* \rm dv$), and signal to noise ratio (S/N) of each transition. In the last column, we also indicate the line blending and the species with which the line is blended. ``Unblended" lines refer to those transitions that are not blended with other known molecular species.}
\tablenotetext{a}{ The S/N is calculated from the integrated signal ($\int T_A ^* \rm dv$) and the noise level $\sigma=\rm rms\times\sqrt{\delta v \times \rm FWHM}$, where $\delta v$ is the velocity resolution of the spectra, and the FWHM is fitted from the data.}
\tablenotetext{b}{Frequency uncertainty to the last digit.}
\tablenotetext{c}{Fitted integrated line intensity uncertainty.}
\label{tab:MgS}
\end{deluxetable*}

\subsection{Tentative detection of calcium oxide (CaO)}

\begin{figure*}
    \centering
\plotone{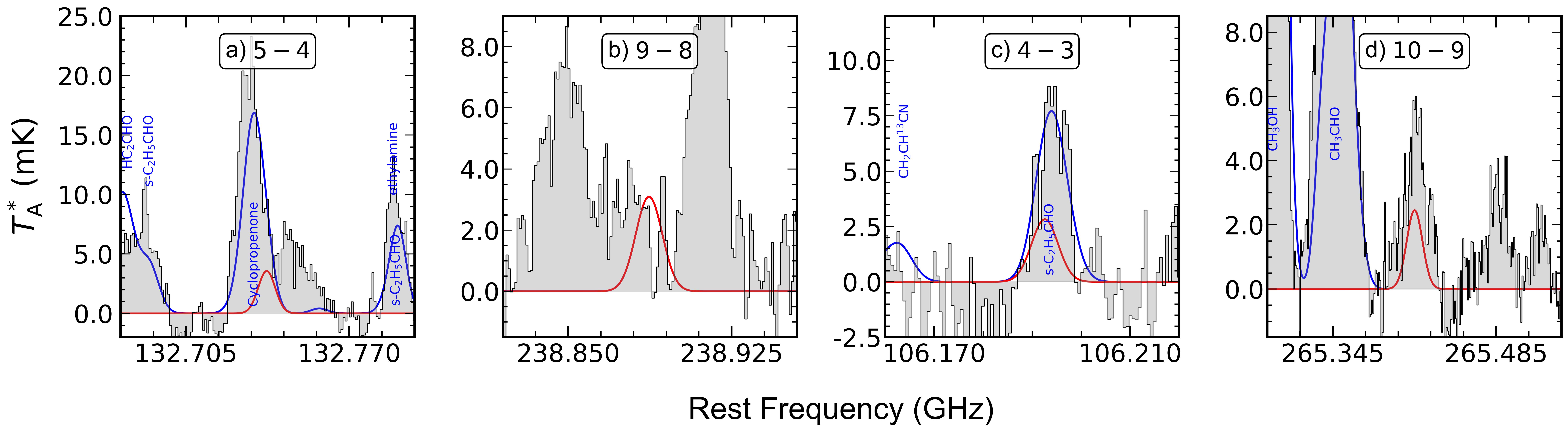}
\caption{Transitions of calcium oxide (CaO) measured towards G+0.693, displayed in order of decreasing line intensity. Two of the observed transitions are unblended while the other two appear heavily blended with other molecular species. The observed spectra are represented as grey histograms. The best LTE fit derived with \textsc{Madcuba} is shown with a red line, while the blue line shows the expected molecular emission from all of the molecular species identified to date in our survey. The quantum numbers representing the J+1$\rightarrow$J  transitions are shown in the upper part of each panel.
\label{fig:CaO}}
\end{figure*}

We have covered seven transitions of CaO in the spectroscopic survey of G+0.693. Only two out of seven transitions observed are unblended while two other are partially blended with transitions from other molecules although the total predicted line profiles reproduce nicely the observed spectra (see Figure~\ref{fig:CaO} and Table~\ref{tab:CaO}). It is worth noting that transition 10-9 at 265.415153 GHz (fourth panel) could be blended with other unidentified species (U-species). Due to the limited number of unblended transitions, although the model of CaO can be fitted to the residual of the emission after subtracting the emission of all other modeled species, it cannot be considered conclusive and therefore the detection is claimed to be only tentative. The rest of transitions covered within the survey lie within the noise and therefore are not shown here. 

To derive the physical parameters of the CaO emission, we follow the same analysis as described in previous sections. As for MgS, we fix the central radial velocity to $v_{\rm LSR}=69.0$ km s$^{-1}$ and the FWHM to 20 km s$^{-1}$, to allow the convergence with \textsc{Slim}-\textsc{Autofit}. The obtained excitation temperature and column density of CaO are $T_{\rm ex}=28.6\pm9.6$ K and $N=(2.0\pm0.5)\times10^{10}$ cm$^{-2}$, respectively (see Table~\ref{tab:PP}). This translates into a molecular abundance of $(1.5\pm0.4)\times10^{-13}$ with respect to H$_2$. We do not show the rotational diagram of CaO because we only have two unblended lines with which to perform this analysis. 

\begin{deluxetable*}{ccccccccc} 
\tablecaption{List of observed transitions of calcium oxide.}
\tablehead{\colhead{Frequency} & \multicolumn{2}{c}{Transition} & \colhead{log I} & \colhead{E$_{\rm u}$}  &  \colhead{rms} & \colhead{$\int T_A ^* \rm dv$} & \colhead{S/N$^a$} & \colhead{Blending}
\\
\colhead{(GHz)} & \multicolumn{2}{c}{($J-J'$ $v^c$)} & \colhead{($\rm
nm^{2}~MHz$)}& \colhead{(K)} & \colhead{(mK)} & \colhead{(mK km $s^{-1}$)} & 
\colhead{} & \colhead{}
}
\startdata 
132.737155 (30)$^b$ & 5-4 & 0 & -1.0466 & 19.0 & 0.9 & 68 (5)$^d$ & 13.9 & Cyclopropenone\\
238.887128 (46)$^b$ & 9-8 & 0 & -0.3326 & 57.1 & 2.4 & 59 (9)$^d$ & 6.4 & Unblended\\
106.192587 (30)$^b$ & 4-3 & 0 & -1.3290 & 12.7 & 1.7 & 54 (10)$^d$ & 5.5 &  CH$_3$OCHO\\
265.415153 (74)$^b$ & 10-9 & 0 & -0.2129 & 69.7 & 0.7 & 47 (3)$^d$ & 17.2 & Blended with U-line\\
\enddata
\tablecomments{We provide the transition frequencies, quantum numbers, base 10 logarithm of the integrated intensities at 300 K (log I), energies of the upper levels (E$_{\rm u}$), noise level (rms), area ($\int T_A ^* \rm dv$), signal to noise ratio (S/N) of each transition. In the last column, we indicate line blending and the species with
which the line is blended. ``Unblended" lines refer to those transitions that are not contaminated by other known molecular species.}
\tablenotetext{a}{The S/N is calculated from the integrated signal ($\int T_A ^* \rm dv$) and the noise level $\sigma=\rm rms\times\sqrt{\delta v \times \rm FWHM}$, where $\delta v$ is the velocity resolution of the spectra, and the FWHM is fitted from the data.}
\tablenotetext{b}{Frequency uncertainty to the last two digits.}
\tablenotetext{c}{$v$ corresponds to the vibrational state.}
\tablenotetext{d}{Fitted integrated line intensity uncertainty.}

\label{tab:CaO}
\end{deluxetable*}


\subsection{Non-detections of other Na-, Mg-, and Ca-bearing molecules}
\label{sec:otherMetals}

Besides NaS, MgS and CaO, we have searched for other Na-, Mg- and Ca-bearing metal molecules for which spectroscopy is available.
None of these species have been detected towards G+0.693, so we have computed their 3$\sigma$ upper limits to their column densities analysing the rms noise level measured for the non-detected brightest transition, and using the same values of $T_{\rm ex}$, $v_{\rm LSR}$ and FWHM inferred for NaS (for the Na-bearing species), for MgS (for the Mg-bearing moecules) and for CaO (for the Ca-bearing species). The measured upper limits to the column densities and abundances can be found in Table~\ref{tab:uplim}, and they can be as low as $\leq$4$\times$10$^{9}$ cm$^{-2}$, which corresponds to an abundance  $\leq$3$\times$10$^{-14}$ with respect to H$_2$. In the Appendix, we present a comparison between these upper limits and the abundances of Na-, Mg- and Ca-bearing species detected toward evolved stars.

\section{Discussion} \label{sec:Discussion}

\subsection{Validity of the detections}
\label{sec:validity}

To check the validity of the NaS and MgS detections, we have followed the criteria of \citet{snyder2005} and \citet{halfen2006} for the definition of an unblended line. The criteria used are the following:

1. The frequency uncertainty of the transition is $\leq$100 kHz (see Tables$\,$\ref{tab:NaS} and \ref{tab:MgS}).

2. The V$_{LSR}$ of the emission of NaS and MgS should lie at 69$\,$km$\,$s$^{-1}$ with a tolerance of $\pm$5$\,$km$\,$s$^{-1}$. The measured V$_{LSR}$ for NaS is 66.5$\,$km$\,$s$^{-1}$, while it is 69$\,$km$\,$s$^{-1}$ for MgS (see Table$\,$\ref{tab:PP}), well within those limits. Note that this V$_{LSR}$ tolerance is significantly narrower than the one assumed by \citet{snyder2005} for Sgr B2(N).

3. The linewidths of the NaS and MgS lines should lie between 15 and 22$\,$km$\,$s$^{-1}$, as observed for other molecular species toward G+0.693 \citep[see e.g.][]{Rivilla2020,Massalkhi2023}. The linewidths that best reproduce the NaS and MgS spectra are 15$\,$km$\,$s$^{-1}$ (Table$\,$\ref{tab:PP}), consistent with these values.
 
4. The unblended lines should not show too great intensity. As shown by the LTE fits of Figures$\,$\ref{fig:NaS} and \ref{fig:MgS} obtained with SLIM-MADCUBA, the observed spectral features are perfectly reproduced by the expected LTE intensity.

5. There must be spectral features at all favorable, physically connected transitions over a sufficiently large wavelength range. As mentioned in Sections$\,$\ref{sec:Nas} and \ref{sec:Mg}, we have detected the full progression of $J$+1$\rightarrow$$J$ transitions of NaS and MgS covered in our broadband spectral survey across (non-continuous) 124 GHz.  

6. \citet{snyder2005} considered that if the observed feature appeared to be a blend of transitions from several species, this line was not used in the analysis. As explained in Section$\,$\ref{sec:Nas}, the observed lines of NaS and MgS are unblended from known molecular species, although some transitions may appear blended with emission from U-species. To determine the level of contribution of the U-species, we have obtained the residuals after subtracting the LTE fit of the NaS and MgS lines, and measured the area of the residuals for the velocity range between $\pm$FWHM/2, with FWHM the measured linewidth of the NaS and MgS lines (in this case, 15$\,$km$\,$s$^{-1}$ for both species). If this contribution is $\leq$25\%, the transition is labeled as ``Unblended$"$ in Tables$\,$\ref{tab:NaS} and \ref{tab:MgS}, and therefore, they are used in our analysis and to claim a detection of these molecular species. If instead, the contribution is $\geq$25\%, the transitions are labeled as ``Blended with U-line$"$ in Tables$\,$\ref{tab:NaS} and \ref{tab:MgS}, and they are not used to claim the detection. This is the case of e.g. the 2$\rightarrow$1 line of MgS or of the 15/2$\rightarrow$13/2 and 11/2$\rightarrow$9/2 lines of NaS.

Following \citet{halfen2006}, we have estimated the confidence level of the NaS and MgS detections considering the ``Unblended$"$ transitions. Assuming the confusion limit \citep[see][]{halfen2006}, the chance of finding a line at the V$_{LSR}$ of our source with a tolerance of $\pm$3$\,$km$\,$s$^{-1}$ (i.e. the difference in V$_{LSR}$ observed for NaS and MgS) is $p$=32\% assuming simple Gaussian profiles and a velocity coverage of the observed feature above the rms noise level equivalent to the FWHM at 1/3 of the peak intensity. The detection of each successive line after the initial one will have a probability $p$(1$^{st}$)$\times$$p$(2$^{nd}$)=$p^2$. Consequently, three unblended lines provides a probability of chance alignment of $p^3$=3.3\%, which implies a confidence level of 96.7\% for the detection of NaS and MgS. If four or five lines were considered (i.e. if we included the transitions blended with U-lines), we would obtain confidence levels of 99\% and 99.7\% for the detection of NaS and MgS, respectively. Note that in any case these are conservative estimates, since we have assumed the confusion limit in our calculations and this limit has not been reached yet in our survey toward G+0.693.


\subsection{Uncertainties in the determination of the abundance of NaS and MgS}
\label{sec:uncertainties}

The derived abundance for NaS and MgS within the beam of our single-dish observations is 3.7$\times$10$^{-13}$ and 4.4$\times$10$^{-13}$, respectively, orders of magnitude lower than the cosmic abundance of Na (1.74$\times$10$^{-6}$) and Mg (3.98$\times$10$^{-5}$; see Section$\,$\ref{sec:Analysis and Results}). One possibility is that the emission of NaS and MgS is not uniformly distributed across the single-dish beam of our observations and hence it is largely diluted.

As shown by \citet{zeng2020}, the SiO (5-4) emission observed with the SMA toward the position of G+0.693, reveals several compact structures with typical sizes of 10$"$. Assuming a source size of 10$"$, the abundance of NaS and MgS would increase to $\sim$10$^{-10}$ for an average single-dish beam of $\sim$30$"$ in our Yebes 40m, IRAM 30m and APEX observations. This abundance would be even higher if the source size were smaller. For instance, if the size of the NaS and MgS emitting region were 2$"$, the abundance of these species would increase to $\sim$3$\times$10$^{-9}$. This implies beam filling factors ranging from $\sim$30 to 6800 for a source size decreasing from 30$"$ to 2$"$. Interferometric observations of the emission of NaS and MgS are needed to constrain their actual abundance in G+0.693 and thus, to infer how much of these metal sulfides reside in dust grains.

\subsection{Correlation between the derived T$_{\rm ex}$ for the metal molecules and their dipole moment.}

In Section \ref{sec:Analysis and Results}, we have found different excitation temperatures for NaS, MgS and CaO. One possible explanation for these differences is that these metal molecules present different dipole moments. 
As shown in  Table~\ref{tab:TexDip}, there is a clear trend for the metal molecules to show lower $T_{\rm ex}$ for increasing dipole moment. Indeed, while the excitation temperature of NaS is $T_{\rm ex}$=17.5 K with $\mu$=9.14 Debye, $T_{\rm ex}$ is 38.5 K for MgS with a lower dipole moment of $\mu$=7.07 D \citep{Chambaud2008}. As mentioned in Section$\,$\ref{sec:Nas}, since these values are lower than the T$_{kin}$ of the gas measured in this cloud \citep[][]{Zeng2018}, the emission from these metal molecules is sub-thermally excited.
For molecules with higher dipole moments, the Einstein $A$-coefficients are much larger, which implies higher critical densities since are proportional to the square of the dipole moment.  Given that the H$_2$ volume density of the gas towards G+0.693 is of a few 10$^4$ cm$^{-3}$ \citep[see e.g.][]{zeng2020}, the H$_2$ density likely lies well below the critical density of the transitions of metal molecules, clearly deviating from LTE. The higher the dipole moment, the higher the critical density thus resulting in a lower derived $T_{\rm ex}$. Note that this trend was already noted for magnesium radicals MgC$_5$N and MgC$_6$H towards the Asymptotic Giant Branch (AGB) star IRC+10216 \citep{Pardo2021}.

\begin{deluxetable}{ccc} 
\tablecaption{Derived $T_{\rm ex}$ for the observed metal molecules and their dipole moment.}
\label{tab:TexDip}
\tablehead{\colhead{Molecule}  & \colhead{$T_{\rm ex}$} & \colhead{$\mu$}
\\
\colhead{} & \colhead{(K)}& \colhead{(Debye)}
}
\startdata 
MgS & $38\pm5$ & 7.07\\
CaO & $29\pm10$ & 8.87\\
NaS & $18\pm4$ & 9.14\\
\enddata
\end{deluxetable}

\subsection{Understanding the origin of NaS, MgS and CaO in G+0.693}
\label{sec:origin}

As mentioned in previous sections, metals are heavily depleted onto dust grains and hence, their detection in the gas phase is challenging \citep{Field1974, Savage1996, Savaglio2003, Jenkins2009, DeCia2016, Roman2021, Konstant2023, Konstantopoulou2024}. The only way to see these metals back into the gas phase is to partially/totally destroy dust grains, especially their silicate cores. The molecular cloud G+0.693 is known to be affected by low-velocity shocks likely produced by a cloud-cloud collision \citep{zeng2020}. The shock chemistry of molecular ions such as PO$^+$ and HOCS$^+$ in G+0.693 has successfully been modelled using the UCLCHEM code\footnote{See https://uclchem.github.io/ and \citet{holdship2017}.} assuming a magneto-hydrodynamic C-type shock with a shock speed v$_s$=20 km s$^{-1}$, a preshock H$_2$ volumn density of 10$^4$ cm$^{-3}$, and an enhanced cosmic-ray ionization rate $\geq$10$^{-15}$ s$^{-1}$ \citep[see][]{Rivilla2022b,Sanz2024}. However, slightly higher shock velocities are needed to sputter the silicate material from grain cores \citep[$\geq$25-30 km s$^{-1}$; see][]{gusdorf2008,jimenez2008}, which suggests that the large-scale shock in G+0.693 could be more energetic than previously thought. The molecular gas in Galactic Center molecular clouds is highly turbulent \citep[linewidths ranging between 20 and 50 km s$^{-1}$;][]{bally1987,lis1998}, and thus this is a plausible scenario. After the sputtering of the grain cores, diatomic metal-bearing molecules could be released directly into the gas phase during this process \citep{jimenez2008}, or they could be produced in the hot post-shocked gas by gas-phase reactions after the release of metals in their atomic form \citep[e.g.][]{gusdorf2008}. In the following, we explore these two possibilities for the formation of NaS, MgS and CaO in G+0.693.

\subsection{Chemistry of MgS}

Magnesium has an ionization potential of 7.64 eV, which implies that this element remains mainly ionized in space even behind large optical depths. Laboratory experiments have shown that due to this fact, MgS dust can easily form by direct nucleation from the gas phase in evolved stars \citep[see][]{kimura2005}. The far-IR band of MgS at $\sim$30 $\mu$m \citep[which is found to be very sensitive to the shape and/or surface of the MgS grains; see][]{kimura2005} has been identified towards carbon-rich AGB stars \citep[][]{goebel1985}, which indicates that MgS is likely part of the refractory core of circumstellar dust that is expelled into the ISM \citep{Begemann1994}. 

Alternatively, MgS in G+0.693 may be generated via gas-phase reactions. The only information available is that its gas-phase formation could occur via the three-body exothermic reaction 2Mg + S$_2$ $\rightarrow$ 2MgS, which presents an energy release of $\sim$38 kJ/mol (\cite{Takano1989}). This reaction, however, does not occur in G+0.693 because its H$_2$ gas volume densities are too low \citep[of some 10$^4$ cm$^{-3}$;][]{zeng2020} for three-body reactions to take place.

Silicates such as olivines and pyroxenes are the dominant compounds of the refractory cores in interstellar dust \citep{Fabian2001, DRAINE2003,min2007} and both contain Si and Mg. We can thus use the chemistry of Si as a proxy for the gas-phase chemistry of Mg. For Si-bearing species, SiO is either released directly from grains in shocks \citep{martin-pintado1992,jimenez2008} or could form in the gas phase after the release of atomic Si into the gas phase \citep[][]{schilke1997,gusdorf2008}. In contrast, SiS is thought to form in the gas-phase\footnote{As explained in \citet{Massalkhi2023}, SiS can be formed throughout multiple reactions: SiH + S and SiH + S$_2$, the reaction S+($^4$S) + SiH$_2$($^1$A$_1$) or the reaction of atomic silicon with either SO or SO$_2$. \citet{Forten2024} also suggests the alternative formation route
SiH+SH → H$_2$ +SiS.} after the passage of a shock \citep[see][]{Massalkhi2023}. The formation of MgS could be due to ion-neutral gas-phase reactions and neutral-neutral gas-phase reactions such as HMgS$^+$ + e$^-$ → H + MgS or Mg + S/S$_2$/HS, respectively \citep{Podio2017}. The big difference between the abundances of SiS and MgS measured towards the G+0.693 molecular cloud (of 3.9$\times$10$^{-10}$ for SiS and of 4.5$\times$10$^{-13}$ for MgS; see \citet{Massalkhi2023} and Section~\ref{sec:Mg}), could be due to the stronger bond of SiS (R$_{\rm Si-\rm S}=1.9$ \AA, with R the inter-atomic distance), which would be harder to destroy that that of MgS (R$_{\rm Mg-\rm S}=2.6$ \AA). However, note that, if the MgS emission were compact, the derived abundance of this metal molecule could be several orders of magnitude higher than reported here (see Section$\,$\ref{sec:uncertainties}).



Indeed, the large difference in abundances between SiS and MgS is unexpected because the cosmic abundance ratio of magnesium and silicon is Mg/Si $\approx$ 1 \citep{Turner1985}. The same applies to SiO and MgO towards G+0.693, for which the measured abundance ratio\footnote{The abundance of SiO towards G+0.693 is 5.3$\times$10$^{-9}$ \citep{Massalkhi2023}, while the upper limit to the abundance of MgO is $\geq$3.2$\times$10$^{-14}$ (see Table~\ref{tab:uplim}).} is SiO/MgO $\geq$ 1.6$\times$10$^5$. Why are Si-bearing molecules more abundant than Mg-bearing ones in the G+0.693 molecular cloud? Following a similar argument as above, a possible explanation is that the stronger SiO bond (R$_{\rm Si-\rm O}=1.6$ \AA; R$_{\rm Mg-\rm O}=2.1$\AA) resists better the sputtering process \citep[see][]{Turner1985}. Furthermore, since MgO likely presents a ion-bonded crystalline structure, it is more difficult to volatize from dust grains by an interstellar shock \citep{Turner1985}.

\subsection{Chemistry of NaS}

To our knowledge, there is no information in the literature about the presence of NaS dust in evolved stars or about the chemistry of NaS in the gas phase. In the case of Na, we know that this element is one of the least depleted metals and that it has a very low ionization potential (5.14 eV), so Na atoms in the gas phase are mostly in its ionized form. However, Na$^+$ is unreactive with many common neutral species which makes it an important source of free electrons in molecular dark clouds. As a result, very few Na-bearing molecules have been detected so far in the ISM. 

To investigate the possible formation routes of NaS we first review the detections of other Na-bearing molecules in order to set a starting point. For NaCl, \citet{Acharyya2024} proposed its formation on dust grains in the presence of a water-ice mantle since the sodium ions can form a strong bond with water ice. A neutral chlorine atom can land on top of the sodium ion to form a weakly bound NaCl$^+$ species, which can react with an electron on the surface yielding the desorption of NaCl. NaS could then be produced in an analogous process with a sulfur atom landing on top of a sodium ion. 
Another possibility is that, similarly to the reaction between Na$^+$ and O$_2$ to form NaO$_2 ^+$ (presumably via an association reaction), 
NaS could form via the reaction between Na and S$_2$ producing NaS$_2 ^+$. NaS would thus be the product of the electron recombination reaction NaS$_2 ^+$ + e$^-$ $\rightarrow$ NaS + S. However, laboratory data and/or quantum chemical calculations are needed to obtain accurate information about the formation route of NaS under interstellar conditions.

\subsection{Chemistry of CaO}

As for NaS, there is no information available in the literature about the chemistry of CaO. The only Ca-bearing molecules that have been detected to date in space are Ca-containing cyanides of the type Ca(C$_{2n+1}$N), with $n$ = 0, 1, 2 \citep{Cernicharo2019a}. These molecules could form in circumstellar envelopes by a two-step process initiated by the radiative association of Ca$^+$ with long cyanopolyynes followed by the dissociative recombination with electrons of the Ca$^+$/NC$_{2n+1}$H complexes \citep{Cernicharo2019a}. CaO may form in a similar fashion through the ion-neutral reactions Ca$^+$ + O$_2$ $\rightarrow$ CaO$_2^+$ or Ca$^+$ + O $\rightarrow$ CaO$^+$, which then react with electrons to form CaO. Alternatively, CaO could be formed in a similar way as proposed for SiO, either on dust grains and released directly from grains \citep{martin-pintado1992}, or through the neutral-neutral reactions Ca + OH $\rightarrow$ CaO + H and Ca + O$_2$ $\rightarrow$ CaO + O \citep{schilke1997}. Once again laboratory data and quantum chemical calculations are needed to understand the formation of these metal-bearing molecules.

\subsection{Differences between O-bearing and S-bearing molecules}

The cosmic abundance ratio of magnesium and silicon is Mg/Si $\approx$ 1 \citep{Turner1985}. From this, one would expect the abundance ratios of MgO/MgS to be similar to those of SiO/SiS towards the G+0.693 molecular cloud. However, while the SiO/SiS abundance ratio towards G+0.693 is 14 \citep{Massalkhi2023}, the MgO/MgS abundance ratio is $<$ 0.06 (see Section~\ref{sec:Mg}). This implies that MgS is formed more efficiently towards G+0.693 than MgO, or that MgO is destroyed more efficiently, which contrasts with the behaviour found for SiO and SiS in the same source. 

\begin{figure}
    \centering
    \includegraphics[width=0.48\textwidth]{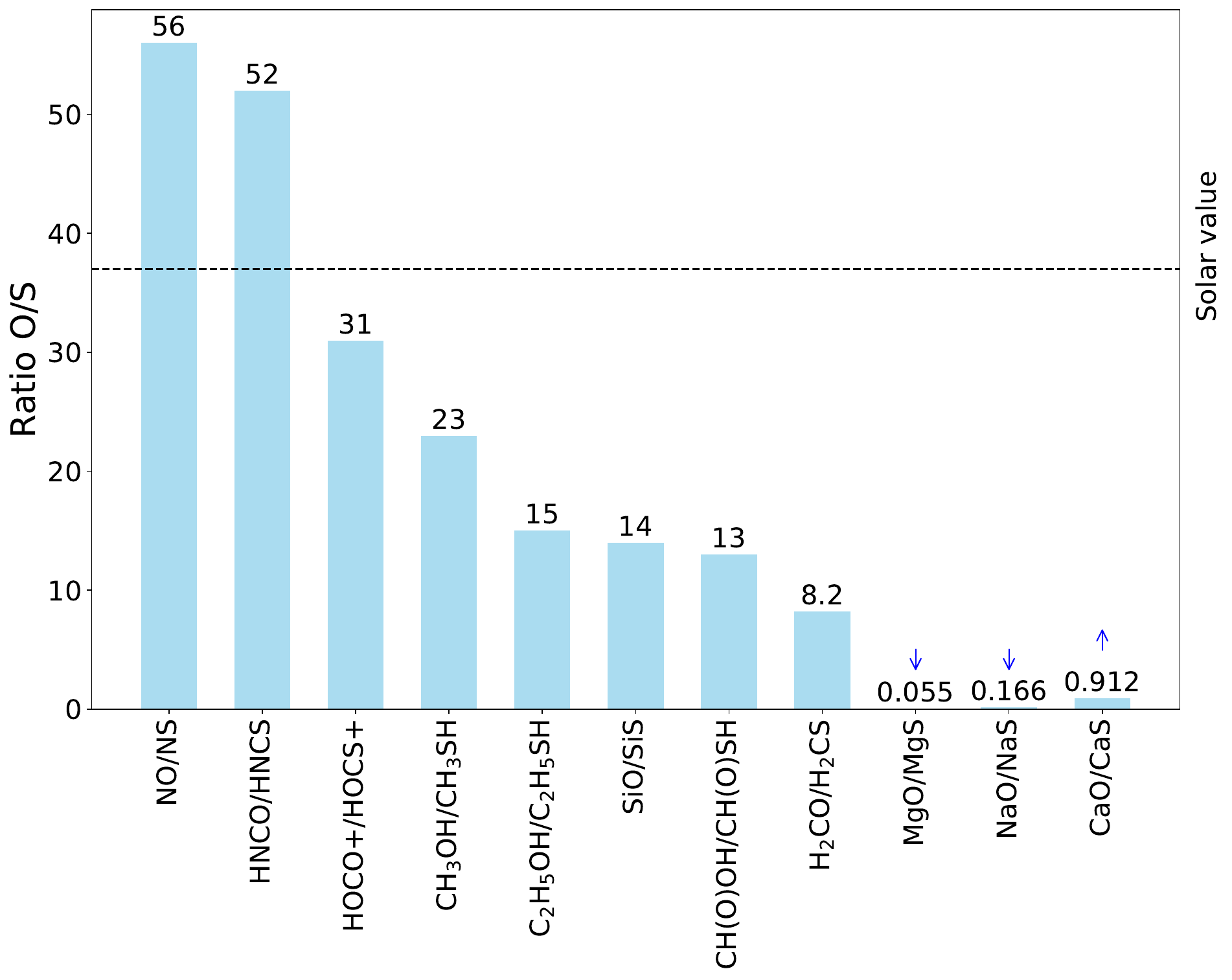}
\caption{Relative O/S ratio of detected molecules towards G+0.693 \citep[CO/CS, H$_2$CO/H$_2$CS][]{Sanz2024}; \citep[NO][]{Rivilla2022a}; \citep[NS][]{Sanz2024b}; \citep[HNCO][]{Zeng2018}; \citep[HNCS][]{Sanz2024}; \citep[HOCO$^+$/HOCS$^+$][]{Sanz2024b}; \citep[SiO/SiS][]{Massalkhi2023}; \citep[CH$_3$OH/CH$_3$SH, C$_2$H$_5$OH/C$_2$H$_5$SH, CH(O)OH/CH(O)SH][]{Rodriguez2021a}. CO/CS=3500 ratio is not included since it is several orders of magnitude above the solar value \citep[see][]{Sanz2024}. The arrow pointing downwards denotes an upper limit and the arrow pointing upwards denotes a lower limit.}
\label{fig:Ratio}
\end{figure}

In Figure~\ref{fig:Ratio}, we present the abundance ratios between metal oxides and metal sulfides and compare them with those obtained not only for SiO/SiS but also for molecules like HNCO and COMs, detected towards G+0.693. This figure strengths the idea that while SiO and the O-bearing non metal-bearing molecules are clearly more abundant than their S-bearing counterparts (by factors $\geq$13), the opposite is true for metals. Indeed, the measured abundance ratios are $\leq$ 0.055 for MgO/MgS, $\leq$ 0.17 for NaO/NaS, and $\geq$ 0.91 for CaO/CaS. The discovery of other S-bearing molecules towards the G+0.693 molecular cloud (as e.g. mono thioformic acid HC(O)SH and HNSO), has been attributed to sulfur being significantly undepleted toward this cloud \citep[][]{Rodriguez2021a, Sanz2024b}. Therefore, G+0.693 represents an excellent astrochemical laboratory to search for new S-bearing molecules in the ISM. 


\section{Summary and Conclusions.} \label{sec:Conc}
We have searched for metal-bearing molecules towards the G+0.693 molecular cloud located in the Galactic center, which is believed to be affected by a cloud-cloud collision that releases many molecules into the gas phase via dust grain sputtering. We report the first detection of NaS with a column density of $N=(5.0\pm1.1)\times10^{10}$ cm$^{-2}$, which translates into a fractional abundance with respect to H$_2$ of $(3.7\pm1.0)\times10^{-13}$. We also present the first detection of MgS whose derived column density is $N=(6.0\pm0.6)\times 10^{10}$ cm$^{-2}$, and whose fractional abundance is $(4.4\pm0.8)\times10^{-13}$. CaO has been tentatively detected with a column density of $N=(2.0\pm0.5)\times10^{10}$ cm$^{-2}$ and a derived fractional abundance of $(1.5\pm0.4)\times10^{-13}$. We note, however, that these abundances should be taken as lower limits since the emission of these metal molecules may be compact. Other Na-, Mg-, and Ca-bearing species were also searched for in this cloud, but these searches did not yield any detection.

By comparing the abundances of O-bearing and S-bearing metal molecules, we find that O-bearing species are less abundant than their sulfur-bearing counterparts, which contrasts with the behaviour found so far for non metal-bearing molecules, COMs and Si-bearing molecules in G+0.693. Very little is known about the formation of these metal-bearing molecules. For MgS, which has been detected in dust, its formation likely occur on dust grains and subsequently sputtered into the gas phase in shocks. For NaS and CaO, no theoretical and experimental information is available for these species. Modelling of the chemistry of metal sulfides and metal oxides needs to be carried out in order to understand the mechanisms responsible for the observed abundances of MgS, NaS and CaO towards G+0.693. 


To conclude, these detections are of great importance since they represent the first interstellar metal-bearing molecules involving sulfur that have been detected to date. Future laboratory experiments and quantum chemical calculations are needed in order to understand how these molecules form and behave in the ISM.

\section{Acknowledgments}
    M. R-M acknowledges funding from a JAE-intro (JAEINT-23-01854) from the Consejo Superior de Investigaciones Cient\'{\i}ficas (CSIC). I.J-.S, J.M.-P., V.M.R, L.C, A.M, A.M-H, A.L-G, D.S.A acknowledge funding from grants No. PID2019-105552RB-C41 and PID2022-136814NB-I00 funded by MICIU/AEI/10.13039/501100011033 and by “ERDF/EU”. 
    V.M. R acknowledges support from project number RYC2020-029387-I funded by MICIU/AEI/10.13039/501100011033 and by "ESF, Investing in your future", and from the Consejo Superior de Investigaciones Cient{\'i}ficas (CSIC) and the Centro de Astrobiolog{\'i}a (CAB) through the project 20225AT015 (Proyectos intramurales especiales del CSIC). 
    A. Megías aknowledges funding from grant PRE2019-091471 funded by MICIU/AEI/10.13039/501100011033 and by 'ESF, Investing in your future'.
    M. S. N. acknowledges a Juan de la Cierva Postdoctoral Fellow proyect JDC2022-048934I, funded by MICIU/AEI/10.13039/501100011033 and by the European Union “NextGenerationEU”/PRTR”.
    P.dV. and B.T. thank the support from MICIU through project PID2019-107115GB-C21. B.T. also thanks the Spanish MICIU for funding support from grants ŁPID2019-106110GB-I00, PID2019-106235GB-I00 and PID2022-137980NB-I00 from the Spanish Ministry of Science and Innovation/State Agency of Research MCIN/AEI/10.13039/501100011033 and by “ERDF A way of making Europe”. 
    A.M.- H. acknowledges funds from Grant MDM-2017-0737 Unidad de Excelencia “María de Maeztu” Centro de Astrobiología (CAB, INTA-CSIC).
    A.L.-G. acknowledges the funds provided by the Consejo Superior de Investigaciones Científicas (CSIC) and the Centro de Astrobiología (CAB) through the project 20225AT015 (Proyectos intramurales especiales del CSIC) and from the Spanish Ministry of Science through the project PID2022-136814NB-I00.

%

\vspace{5mm}
\facilities{IRAM:30m, Yebes:40m, APEX}


\software{Madrid Data Cube Analysis (MADCUBA) on ImageJ is a software developed at the Center of Astrobiology (CAB) in Madrid (\cite{Martin2019}).}



\appendix

\section{Comparison with evolved stars.}
\label{sec:ap}

\begin{figure}
    \centering
    \includegraphics[width=0.48\textwidth]{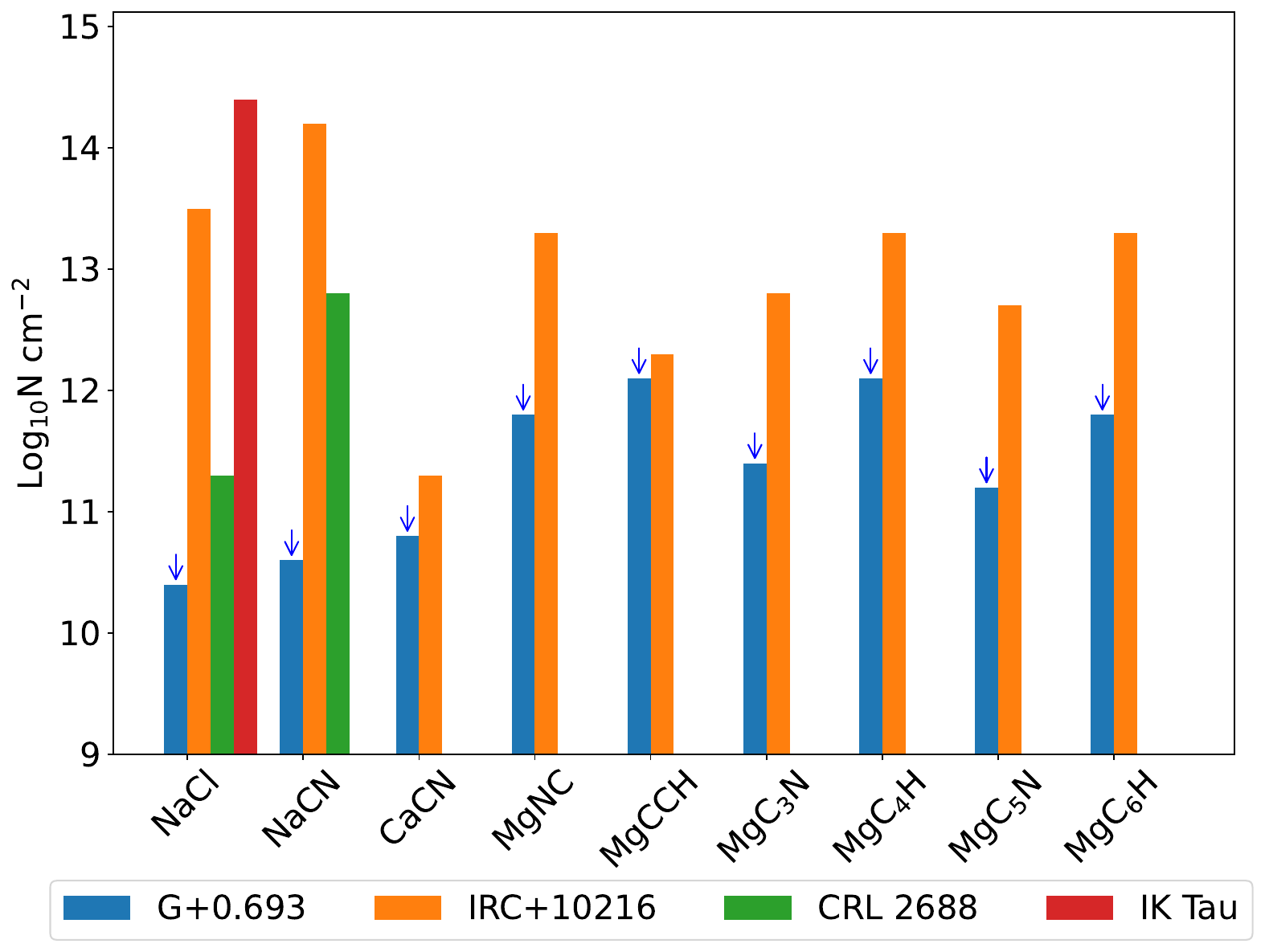}
\caption{Column densities of metal-bearing molecules measured towards different astronomical sources \citep{CernicharoGuelin1987, Turner1994, Highberger2001, Highberger2003a, Highberger2003b, Milam2007, Agundez2014, Cernicharo2019a, Cernicharo2019b, Pardo2021}. NaCl has also been observed towards Orion Source I and the IRAS 16547 binary system \citep[see Section~\ref{sec:intro}][]{Ginsburg19,Tachibana19,Tanaka2021}, but no information is provided in these works about the derived column densities. Arrows denote upper limits.}
\label{fig:AS}
\end{figure}

NaS, MgS and CaO have not been detected toward any other astronomical source. However, other metal-bearing molecules such as NaCl, MgNC or CaCN have been observed towards evolved stars and massive hot disks. Since we have estimated the upper limits to the abundance of these molecules towards G+0.693, we can compare these values with those obtained towards other objects (see Figure~\ref{fig:AS}). From this figure, we find that the upper limits obtained for G+0.693 are typically between one to three orders of magnitude smaller than those measured in evolved stars. 
This could be a consequence of the partial release of material from dust grains by the sputtering produced in the G+0.693 shock (see Section$\,$\ref{sec:origin}).
This suggests that these molecules, once formed in evolved stars in the gas phase, are incorporated into dust grains and injected into the ISM \citep{CernicharoGuelin1987, Kawaguchi1993, Milam2007, Kaminski2013, Cernicharo2019a}. 


\begin{deluxetable*}{cccc} 
\tablecaption{Molecular entries}
\label{tab:spec}
\tablehead{\colhead{Name}  & \colhead{Formula} & \colhead{Molecular catalog entry} & \colhead{References}
}
\startdata 
Sodium sulfide & NaS & CDMS 55508 & \citet{Li1997}\\
Sodium hidrosulfide & NaSH & CDMS 56522 & \cite{Kagi1997}\\
Sodium acetylide & NaCCH & CDMS 48505 & \cite{Brewster1999}\\
Sodium methylidyne & NaCH & CDMS 36501 & \cite{Xin1999}\\
Sodium cyanide & NaCN & CDMS 49510 & \cite{halfen2011}\\
Sodium chloride & NaCl & CDMS 58502 & \cite{caris2002}\\
Sodium fluoride & NaF & CDMS 42502 & \cite{bauer1963}; \cite{veazey1965}\\
Sodium hydroxide & NaOH & CDMS 40509 & \cite{Pearson1973}\\
Sodium oxide & NaO & JPL 39005 &  \cite{Yamada1989}\\
Sodium carbide & NaC & CDMS 35501 & \cite{Sheridan2002}\\
Magnesium sulfide & MgS & JPL 56009 & \cite{Takano1989}\\
Magnesium hidrosulfide & MgSH & CDMS 57516 & \cite{Taleb2001}\\
Magnesium acetylide & MgCCH & CDMS 49507 & \cite{Brewster1999}\\
Magnesium monodiacetylide & MgC$_4$H & CDMS 73505 & \cite{Cernicharo2019b}\\
Magnesium monocyanoacetylide & MgC$_3$N & CDMS 74518 & \cite{Cernicharo2019b}\\
Magnesium monocyanodiacetylide & MgC$_5$N & CDMS 98502 & \cite{Pardo2021}\\
Magnesium monotriacetylide & MgC$_6$H & CDMS 97503 & \cite{Pardo2021}\\
Magnesium monoisocyanide & MgCN & JPL 50009 & \cite{Anderson1994}\\
Magnesium  isocyanide & MgNC & JPL 50010 & \cite{Kawaguchi1993}\\
Magnesium chloride & MgCl & CDMS 59501 & \cite{Bogey1989}\\
Magnesium fluoride & MgF & CDMS 43503 & \cite{Anderson1994}\\
Magnesium monohydroxide & MgOH & CDMS 41508 & \cite{barclay1992}\\
Magnesium oxide & MgO & JPL 40007 & \cite{kagi2006}\\
Calcium oxide & CaO & CDMS 56515 & \cite{creswell1977}\\
Calcium sulfide & CaS & JPL72001 & \cite{Takano1989}\\
Calcium hidrosulfide & CaSH & CDMS 73504 & \cite{Taleb1996}\\
Calcium monoacetylide & CaCCH & CDMS 65511 & \cite{amderson1995}\\
Calcium monoisocyanide & CaNC & CDMS 66503 & \cite{steimle1993}; \cite{scurlock1994}\\
Calcium carbide & CaC & CDMS 52505 & \cite{halfen2002}\\
Calcium monomethyl& CaCH & JPL 55002 & \cite{Anderson1996}\\
Calcium chloride & CaCl$_3$ & CDMS 75502 & \cite{Moller1982}\\
Calcium fluoride & CaF & CDMS 59502 & \cite{anderson21994}\\
Calcium monohydroxide & CaOH & CDMS 57501 & \cite{Scurlock1993}; \cite{Ziurys1992}\\
Calcium deuteride & CaD & JPL 42004 & \cite{Frum1993}\\
\enddata

\label{tab:TexDip}
\end{deluxetable*}

\begin{deluxetable}{lcc}
\tabletypesize{\scriptsize}
\tablewidth{\textwidth} 
\tablecaption{Upper limits to the column density and the abundance for the Na-, Mg- and Ca-bearing molecules measured towards G+0.693. \label{tab:uplim}}
\tablehead{
\colhead{Molecule} & \colhead{$N$}& \colhead{Abundance$^a$}\\
\colhead{} & \colhead{($10^{10}$ cm$^{-2}$)}& \colhead{($10^{-13}$)}\\}
\startdata 
NaSH & $<$4.9 & $<$3.6\\
NaCCH & $<$3.2 & $<$2.4\\
NaCH & $<$1.7 & $<$1.3\\
NaCN & $<$4.4 & $<$3.6\\
NaCl & $<$2.6 & $<$1.9\\
NaF & $<$7.4 & $<$5.5\\
NaOH & $<$8.9 & $<$6.6\\
NaO & $<$1.3 & $<$9.8\\
NaC & $<$4.9 & $<$3.6\\
\hline
Mg$^{34}$S & $<$1.2 & $<$0.87\\
MgSH & $<$56 & $<$42\\
MgCCH & $<$110 & $<$85\\
MgC$_4$H & $<$110 & $<$85\\
MgC$_3$N & $<$22 & $<$17\\
MgC$_5$N & $<$15 & $<$11\\
MgC$_6$H & $<$56 & $<$42\\
MgCN & $<$110 & $<$83\\
MgNC & $<$56 & $<$42\\
MgCl & $<$16 & $<$12\\
MgF & $<$5.6 & $<$4.2\\
MgOH & $<$23 & $<$17\\
MgO & $<$0.44 & $<$0.32\\
\hline
CaS & $<$2.2 & $<$1.6\\
CaSH & $<$14 & $<$10\\
CaCCH & $<$13 & $<$9.3\\
CaNC & $<$6.6 & $<$4.9\\
CaC & $<$6.5 & $<$4.8\\
CaCH$_3$ & $<$110 & $<$85\\
CaCl & $<$19 & $<$14\\
CaF & $<$14 & $<$10\\
CaOH & $<$23 & $<$17\\
CaD & $<$12 & $<$9.1\\
\enddata
\tablenotetext{a}{The molecular abundance with respect to molecular hydrogen is derived assuming $N_{\rm H_2}=1.35\times10^{23}$ cm$^{-2}$ (\cite{Martin2008}).}
\end{deluxetable}

\bibliography{sample631}{}
\bibliographystyle{aasjournal}



\end{document}